\documentclass{amsart}
\usepackage{natbib,amsmath,amsfonts, xfrac, graphicx, url}
\usepackage[utf8]{inputenc}
\usepackage[T1]{fontenc}
\newtheorem{alg}{Algorithm}
\DeclareMathOperator{\conv}{CH}
\DeclareMathOperator{\lhull}{LH}
\DeclareMathOperator{\uhull}{UH}
\newcommand{\st}{:}
\newcommand{\NN}{\mathbb{N}}
\newcommand{\RR}{\mathbb{R}}
\newcommand{\CL}{\mathcal{L}}
\newcommand{\CO}{\mathcal{O}}
\newcommand{\important}[1]{\textbf{#1}}
\newcommand{\term}[1]{\emph{#1}}
\begin{document}
%\begin{frontmatter}
\title[A note about \emph{``Faster algorithms\dots''}]{A note about \emph{``Faster algorithms for computing Hong's bound on absolute positiveness''} by K.~Mehlhorn and S.~Ray}
\author{Przemysław Koprowski}
\begin{abstract}
We show that a linear-time algorithm for computing Hong's bound for positive roots of a univariate polynomial, described by K.~Mehlhorn and S.~Ray in an article ``Faster algorithms for computing Hong's bound on absolute positiveness'', is incorrect. We present a corrected version.
\end{abstract}
%\end{frontmatter}
\maketitle

\section{Introduction}
Computing an upper bound for real roots of a polynomial is an important problem in computational algebra. It has numerous applications (for instance root separation, to mention just one). Thus, in recent decades there has been intensive effort to find such bounds. Given a univariate polynomial $A = a_0 + a_1x + \dotsb + a_nx^n\in \RR[x]$ with a positive leading coefficient, a good bound for its positive roots is the one obtained in \cite{Hong98}:
\[
H(A) := 2\cdot \max_{\substack{j < n\\ a_j < 0}} \min_{\substack{i > j\\ a_i > 0}}
\left( \frac{-a_j}{a_i} \right)^{\frac{1}{i-j}}.
\]
A na\"{\i}ve, straightforward implementation of this bound clearly has a time complexity of $\CO(n^2)$. \cite{MR10} proposed a geometric approach to computing this bound and presented a very smart algorithm that runs in linear time (with respect to the degree of a polynomial). Unfortunately their algorithm turns out to be incorrect.

The aim of this note is to explain the source of errors as well as to propose a possible corrections so that the resulting algorithm accurately computes the Hong's bound and still works in linear time.

For the sake of clarity, in this paper we adopt the notation of \cite{MR10}, much to a discomfort of the author, who is more used to a different notational convention. Thus, with every nonzero term $a_ix^i$ of a polynomial $A = a_0 + a_1x + \dotsb + a_nx^n$, we associate a pair $\bigl( i, -\lg \lvert a_i\rvert\bigr)$, where $\lg$ denotes a logarithm of base~$2$. We interpret these pairs as points in the real plane~$\RR^2$. A point $p = \bigl( i, -\lg\lvert a_i\rvert\bigr)$ is said to be \term{positive} if $a_i > 0$, otherwise it is called \term{negative}. The set of positive points with absciss\ae{} greater or equal to some fixed integer~$j$ is denoted
\[
P_j^+ := \Bigl\{ \bigl( i, -\lg\lvert a_i\rvert\bigr)\st i > j \text{ and } a_i > 0 \Bigr\}.
\]
The absciss\ae{} of the points~$p_i$ for $i\in \{0, \dotsc, n\}$, being exponents of~$x$ in the original polynomial, are non-negative integers. Thus, if~$p_i$ is a positive point, then~$p_i$ and~$p_n$ are respectively the left-most and right-most points of the convex hull $\conv(P_i^+)$ of~$P_i^+$. Consequently, they split the boundary of $\conv(P_i^+)$ into two chains of points: the \term{lower hull} $\lhull(P_i^+)$ and the \term{upper hull} $\uhull(P_i^+)$. If~$p$ is a point positioned to the left of~$P_i^+$, then there is a unique line that passes through~$p$ and at least one point of~$P_i^+$ and such that all points of~$P_i^+$ are on or above it. This line is called the \term{lower tangent} and denoted $\tau(p, P_i^+)$. The left-most point of the intersection $\tau(p, P_i^+)\cap P_i^+$ is called the \term{point of tangency} of $\tau(p, P_i^+)$. The algorithm of Mehlhorn and Ray keeps track of the following data during execution:
\begin{itemize}
\item $\sigma_i$ is the maximal slope of lower tangents computed so far;
\item $\ell_i$ is the lower tangent to $\conv(P_i^+)$ with the slope~$\sigma_i$;
\item $t_i$ is the point of tangency of~$\ell_i$;
\item $\CL_i$ is the lower hull of~$P_i^+$.
\end{itemize}
In what follows we preserve this notation.

\section{Problems}
The algorithm presented in \cite[Section~3]{MR10} suffers from two interconnected problems:

\subsubsection*{Issue 1.} 
The variable~$t_i$---that stores the point of tangency of the lower tangent with a maximal slope---is initialized at the beginning of the algorithm \cite[Algorithm~1, line~2]{MR10} by setting $t_n := p_n$, but it is \important{never lowered}. Hence it stays at~$p_n$ throughout the whole process. Indeed, analyzing the algorithm, we see that:
\begin{itemize}
\item in line~15, $t_i$ is being passed over by setting $t_i := t_{i+1}$;
\item for $a_i > 0$, the variable~$t_i$ is not set in the pseudo-code, but according to the description \cite[page~680, line $-4$]{MR10} it is again set $t_i := t_{i+1}$;
\item in line~10, a new~$t_i$ is searched to the \important{right} of~$t_{i+1}$.
\end{itemize}

\subsubsection*{Issue 2.}
\begin{figure}
\begin{center}
\includegraphics{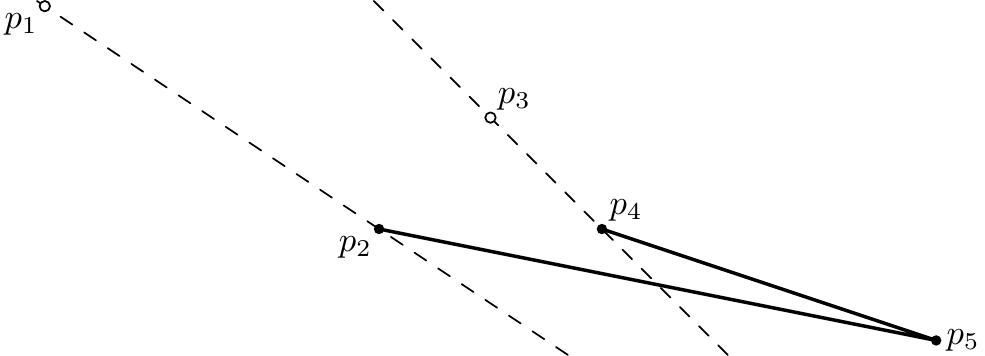}
\caption{\label{fig_points}Configuration of points corresponding to the polynomial $A_{-1} := -1 + 4x^3 - 2x^4 + 4x^5 + 8x^8$. Positive points are marked by filled dots, negative ones by empty dots. Thick lines correspond to lower hulls, dashed lines to lower tangents.}
\end{center}
\end{figure}

The claim \cite[page~680, line $-10$]{MR10}:
\begin{quotation}
\emph{``the tangent point~$t_i$ of $\ell_i = \tau(p_i, P_i^+)$ cannot lie to the left of~$t_{i+1}$''}
\end{quotation}
is false. It would be true if the lower hull of positive points has not changed since we found~$t_{i+1}$. But not without this assumption. If the lower hull has changed, looking for the point of tangency, we must scan the entire lower hull starting from its beginning, not from the previously obtained point~$t_{i+1}$. We show an explicit example. Take a polynomial:
\[
A_\alpha := \alpha + 4x^3 - 2x^4 + 4x^5 + 8x^8,
\qquad\text{with}\quad \alpha < 0.
\]
For $\alpha = -1$, the polynomial has two negative coefficients and so the Hong's bound equals twice the maximum of two minimums:
\begin{align*}
\text{for }a_0 &: 
  \min_{\substack{i > 0\\ a_i > 0}} \left( \frac{-\alpha}{a_i} \right)^{\frac1i} 
  = \min \left\{ \sqrt[3]{\frac14}, \sqrt[5]{\frac14}, \sqrt[8]{\frac18} \right\} 
  = \sqrt[3]{\frac14}\\
\text{for }a_4 &:
  \min_{\substack{i > 4\\ a_i > 0}} \left( \frac{2}{a_i} \right)^{\frac1{i-4}} 
  = \min \left\{ \frac12, \sqrt[4]{\frac14} \right\} 
  = \frac12
\end{align*}
Thus, the maximum is reached for a pair of coefficients~$(a_0, a_3)$ and 
\[
H(A) = 2\cdot \sqrt[3]{\sfrac14} = \sqrt[3]{2}\approx 1.2599. 
\]
Let us analyze what the algorithm of Mehlhorn and Ray does in this case. Ignore for a moment issue~1. The coefficients of~$A$ correspond to points (see Figure~\ref{fig_points}):
\[
p_1 = (0,0),\quad p_2 = (3,-2),\quad p_3 = (4,-1),\quad p_4 = (5,-2),\quad p_5 = (8,-3).
\]
The lower hull $\CL_4 = \lhull(P_4^+)$ of $P_4^+ = \{ p_4, p_5\}$ is just a line segments connecting~$p_4$ with~$p_5$. The lower tangent $\tau(p_3, \CL_4)$ to the convex hull of~$P_4^+$ and passing through~$p_3$ is a line~$\ell_3$ with a slope $s_3 = \frac{(-1) - (-2)}{4 - 5} = -1$. The point of tangency of~$\ell_4$ is $t_3 = p_4$. (If we did not ignore issue~1 here, the algorithm would incorrectly pick up $t_3 = p_5$ and so~$s_3$ would be $\sfrac{-1}{2}$, which is evidently not minimal.) The point~$p_3$ is negative, thus the lower hull of positive points stays intact and we have $\CL_3 = \CL_4$.

The next point to consider is~$p_2$. It is positive, hence the algorithm updates the lower hull. Now, $\CL_2$ is a line segment with endpoints~$p_2$ and~$p_5$. The point~$p_4$ is removed from the lower hull. 

Finally, we consider the point~$p_1$. It is negative again. The lower tangent $\tau(p_1, \CL_2)$ is a line~$\ell_1$ that passes through the newly added point~$p_2$. It has a slope of~$\sfrac{-2}{3}$. It is now evident that in order to find the tangent point, we must scan the lower hull \important{from the beginning}, not from the point $t_3 = p_4$. Not only the point~$p_4$ was removed from the lower hull, but scanning the points to the right of it would result in picking up a point~$p_5$ and the resulting line would have a slope~$\sfrac{-3}{8}$. All in all, the algorithm computes $\max\{ -1, \sfrac{-3}{8} \} = \sfrac{-3}{8}$ and returns $H(A) = 2\cdot 2^{\sfrac{-3}{8}}\approx 1.5422$, which is not the correct Hong's bound for~$A$.

It should be noted that, if we take $\alpha = -8$ in the above polynomial, then the Hong's bound is obtained not from the pair $(p_1, p_2)$ but from $(p_1, p_5)$. This shows that if the lower hull of the set of positive points is rebuilt, we cannon a priori exclude any of the points. All the point of the new lower hull must be scanned to obtain the lower tangent. 

\section{Corrections}
The algorithm of Mehlhorn and Ray can be simply corrected to evade the pitfalls mention in the previous section. The most direct approach is to update the point of tangency~$t_i$, when the lower hull of the positive points changes. A straightforward, but rather na\"\i{}ve, solution is to set~$t_i$ to the first point of~$\CL_i$. The corresponding part of the algorithm would then read:
\emph{\begin{itemize}
\item if~$p_i$ is a positive point \textup(i.e. $a_i > 0$\textup), then:
  \begin{itemize}
  \item compute the lower hull~$\CL_i$ of~$P_i^+$ as described in \cite[Section~2]{MR10};
  \item set $t_i := p_i$, $\sigma_i := \sigma_{i+1}$ and let~$\ell_i$ be a line through~$t_i$ of a slope equal~$\sigma_i$.
  \end{itemize}
\end{itemize}}
Unfortunately, these modifications blow the time complexity of the algorithm up to $\CO(n^2)$, which is the time complexity of the most direct evaluation of Hong's bound. 

One may observe that if a positive point~$p_i$ lies on or above the line~$\ell_{i+1}$, then it will never become a tangent point and consequently the point~$t_i$ may be left intact in this case. Indeed, suppose that~$p_i$ is on or above~$\ell_{i+1}$ and take a subsequent negative point~$p_k$, $k < i$. As explained in \cite[Section~3.2]{MR10}, without loss of generality, we may assume that~$p_k$ lies below~$\ell_{i+1}$, since otherwise it would not contribute to the Hong's bound. But now~$p_k$ is under and~$p_i$ above a previously computed lower tangent~$\ell_{i+1}$. Therefore, a line through~$p_k$ and~$p_i$ is definitely not the lower tangent $\tau(p_k, P_{k+1}^+)$. Thus we may improve upon the previous solution treating positive points as follows (this code replaces \cite[Algorithm~1, lines 19--20]{MR10}):
\emph{\begin{itemize}
\item if~$p_i$ is a positive point \textup(i.e. $a_i > 0$\textup), then:
  \begin{itemize}
  \item let~$t$ be the tangent point of $\tau(p_i, P_{i+1}^+)$, computed by scanning the lower hull~$\CL_{i+1}$ starting from the front;
  \item replace the points before~$t$ in $\CL_{i+1}$ by $p_i$ to obtain~$\CL_i$;
  \item if~$p_i$ lies below the line~$\ell_{i+1}$ then set $t_i := p_i$, $\sigma_i := \sigma_{i+1}$ and let~$\ell_i$ be a line parallel to~$\ell_{i+1}$ and passing through~$p_i$;
  \item otherwise, set $t_i := t_{i+1}$, $\sigma_i := \sigma_{i+1}$ and~$\ell_i := \ell_{i+1}$.
  \end{itemize}
\end{itemize}}

Regrettably even the improved version still has a quadratic time-complexity. In order to recover the linear time complexity, we need to reshape the algorithm more seriously and separate the phase of building of the lower hulls from computation of the lower tangent of a maximal slope. The algorithm we are going to present is a two-pass process. The first pass scans the points in the decreasing order of indices (i.e. right-to-left) and is used to build and store all the lower hulls. Subsequently, in the second pass the algorithm goes through the points left-to-right and computes the sought Hong's bound. 

As said, we are going to store all intermediate lower hulls~$\lhull(P_i^+)$. If we did it na\"\i{}vely and store them as a list of lists, then we would end up with a space complexity of $\CO(n^2)$. Reading and writing this data would need $\CO(n^2)$ time. We can evade this trap, observing that if~$p_i$ is a positive point, then $\lhull(P_i^+)$ is a chain of the form $(p_i, p_j, \text{further points})$ and its ``tail'' $(p_j, \text{further points})$ is a lower hull of~$P_j^+$. Consequently, we may store all the lower hulls as a single array of indices, call it~$V$. For a positive point~$p_i$ we set $V[i] = j$ to be the index of the \important{second} point (the first one is of course~$p_i$) of the lower hull $\lhull(P_i^+)$. Hence the lower hull of~$P_i^+$ equals
\[
\lhull(P_i^+) = \bigl( p_i, p_{V[i]}, p_{V[V[i]]}, \dotsc \bigr).
\]
On the other hand, for a negative point~$p_i$, we let $V[i]$ be the index of the \important{first} point of $\lhull(P_i^+)$, or in other words the index of the first positive coefficient of~$A$ to the right of~$a_i$.

We are now ready to present a corrected algorithm that computes the Hong's bound of a polynomial in linear time.

\begin{alg}
Given a polynomial $A = a_0 + a_1x + \dotsb + a_nx^n$ with a positive leading coefficient, this algorithm computes its Hong's upper bound for positive roots. In what follows, we denote $p_i := \bigl( i, -\lg \lvert a_i\rvert\bigr)$ for $i \in \{0, \dotsc, n\}$.
\paragraph{// First pass: construction of all lower hulls}
\newcounter{storeenumi}
\begin{enumerate}
\item Set $k := n$, it will store the index of the last visited positive point, and initialize $V[n] := -1$;
\item iterate over the coefficients of~$A$ in decreasing order of indices $i\in \{n-1, n-2, \dotsc, 0\}$;
  \begin{enumerate}
  \item if $a_i < 0$, then the lower hull $\lhull(P_i^+)$ does not change, hence set $V[i] := k$ and reiterate the main loop;
  \item if $a_i > 0$, then
    \begin{enumerate}
    \item scan the lower hull 
    \[
    \CL_{i+1} = \lhull(P_{i+1}^+) = \bigl( p_k, p_{V[k]}, p_{V[V[k]]}, \dotsc \bigr)
    \]
    to find the point of tangency of $\tau(p_i, P_{i+1}^+)$ as explained in Section~2 of \cite{MR10};
    \item let $j\in \NN$ be the abscissa of the point found in the previous step, set $V[i] := j$;
    \item update $k := i$;
    \end{enumerate}
  \end{enumerate}
\setcounter{storeenumi}{\theenumi}
\end{enumerate}
\paragraph{// Second pass: computation of the Hong's bound}
\begin{enumerate}
\setcounter{enumi}{\thestoreenumi}
\item let $j := \min\{ i\st a_i < 0\}$ be the lowest index of a negative coefficient of~$A$;
\item scan the lower hull $\lhull(P_j^+) = \bigl( p_{V[j]}, p_{V[V[j]]}, \dotsc \bigr)$ to find the tangent point $p_k$ of $\tau(p_j, P_j^+)$, set $t_j := p_k$, $\ell_j := \tau(p_j, P_j^+)$ and let~$\sigma_j$ be the slope of~$\ell_j$;
\item iterate over the coefficients of~$A$ in an increasing order of indices starting from~$j+1$:
  \begin{enumerate}
  \item if $a_i > 0$, then:
    \begin{enumerate}
    \item set $\sigma_i := \sigma_{i-1}$;
    \item if $p_i = t_i$, then this point gets removed from the lower hull in the next step, hence set $t_i$ to the first point of $\lhull(P_{i+1}^+)$, which is either $p_{i+1}$ if $p_{i+1}$ is positive or $p_{V[i+1]}$ if $p_{i+1}$ is negative, and let~$\ell_i$ be a line parallel to~$\ell_{i-1}$ but passing through~$t_i$;
    \item otherwise pass over: $\ell_i := \ell_{i-1}$ and $t_i := t_{i-1}$;
    \end{enumerate}
  \item if $a_i < 0$, then :
    \begin{enumerate}
    \item if~$p_i$ lies on or above~$\ell_{i-1}$, then ignore it, setting $\ell_i := \ell_{i-1}$, $\sigma_i := \sigma_{i-1}$ and $t_i := t_{i-1}$, reiterate the main loop;
    \item otherwise, when~$p_i$ lies below~$\ell_{i-1}$, the let~$k$ be the abscissa of~$t_i$ and scan the ``tail'' of the lower hull $\lhull(P_i^+)$ consisting of points $(p_{V[k]}, p_{V[V[k]]}, \dotsc )$ to find the tangent point~$p_m$ of $\tau(p_i, P_i^+)$;
    \item if the slope of $\tau(p_i, P_i^+)$ is greater than~$\sigma_{i-1}$, set $t_i := p_m$, $\ell_i := \tau(p_i, P_i^+)$ and $\sigma_i$ the slope of~$\ell_i$;
    \end{enumerate}
  \end{enumerate}
\item return $H(A) = 2^{1 + \sigma_n}$.
\end{enumerate}
\end{alg}

\begin{figure}
\begin{center}
\includegraphics{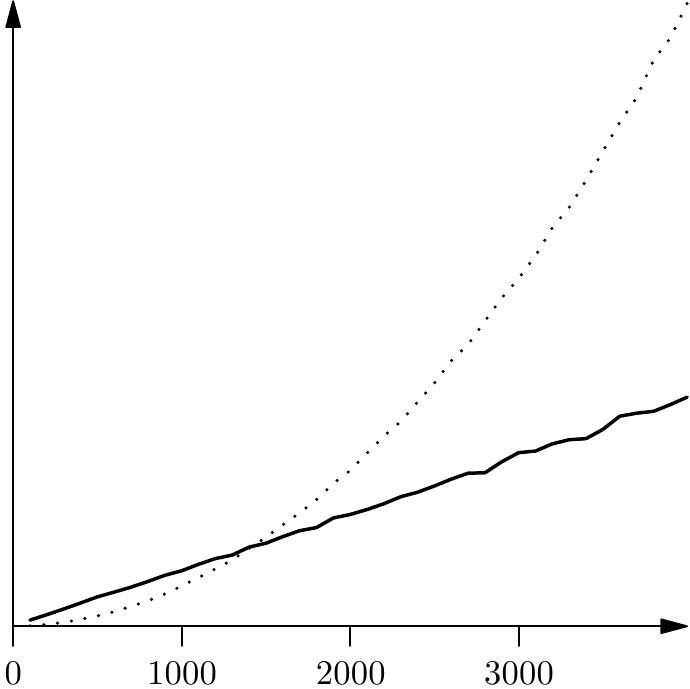}
\caption{\label{fig_graph}Running time of a Sage implementation of a proposed algorithm (solid line) agains a direct implementation (dotted line). The horizontal axis represents degrees of random polynomials.}
\end{center}
\end{figure}

The corrected algorithm was implemented in a computer algebra system \cite{sage}, both to test it correctness and evaluate its speed. The code can be downloaded from authors home page at \url{http://z2.math.us.edu.pl/perry/papersen.html}. It was executed on randomly generated polynomials of varying degrees and the average computation times were compared with running times of a direct implementation of the Hong's bound. Figure~\ref{fig_graph} and Table~\ref{tab_times} summarize the results.

\begin{table}
\begin{center}
\caption{\label{tab_times}Running time of a Sage implementation of a proposed linear-time algorithm (middle column) against a direct quadratic-time implementation (right column). The times are in second for 10 polynomials}
\begin{tabular}{r|r@{.}lr@{.}l}
\multicolumn{1}{c|}{degree} & \multicolumn{2}{c}{linear} & \multicolumn{2}{c}{quadratic}\\\hline
    5 &   0 & 053 &   0 & 003 \\
   10 &   0 & 127 &   0 & 005 \\
   20 &   0 & 268 &   0 & 012 \\
   50 &   0 & 711 &   0 & 038 \\
  100 &   1 & 492 &   0 & 139 \\
  200 &   2 & 979 &   0 & 473 \\
  500 &   7 & 382 &   2 & 625 \\
 1000 &  13 & 821 &  10 & 279 \\
 2000 &  28 & 116 &  40 & 378 \\
 5000 &  72 & 341 & 245 & 940 \\
10000 & 148 & 879 & 991 & 194 \\
\end{tabular}
\end{center}
\end{table}

% \bibliographystyle{elsart-harv} 
% \bibliography{note}

\begin{thebibliography}{3}
\expandafter\ifx\csname natexlab\endcsname\relax\def\natexlab#1{#1}\fi
\expandafter\ifx\csname url\endcsname\relax
  \def\url#1{\texttt{#1}}\fi
\expandafter\ifx\csname urlprefix\endcsname\relax\def\urlprefix{URL }\fi

\bibitem[{Hong(1998)}]{Hong98}
Hong, H., 1998. Bounds for absolute positiveness of multivariate polynomials.
  J. Symbolic Comput. 25~(5), 571--585.
\newline\urlprefix\url{http://dx.doi.org/10.1006/jsco.1997.0189}

\bibitem[{Mehlhorn and Ray(2010)}]{MR10}
Mehlhorn, K., Ray, S., 2010. Faster algorithms for computing {H}ong's bound on
  absolute positiveness. J. Symbolic Comput. 45~(6), 677--683.
\newline\urlprefix\url{http://dx.doi.org/10.1016/j.jsc.2010.02.002}

\bibitem[{Sage(2016)}]{sage}
Sage, 2016. {S}age {M}athematics {S}oftware {S}ystem ({V}ersion 7.3).
\newline\urlprefix\url{\url{http://www.sagemath.org}}

\end{thebibliography}

\end{document}